# Single Molecule Ferroelectrics via Conformational Inversion: An Electronic Structure Investigation

*Xinfeng Quan[1], Geoffrey R. Hutchison[2*]*

[1]Sichuan University-Pittsburgh Institute, Chengdu, Sichuan, CN 610065

[2]Department of Chemistry, University of Pittsburgh, 219 Parkman Avenue,
Pittsburgh, Pennsylvania 15260, USA

* Corresponding author: geoffh@pitt.edu

ABSTRACT: Ferroelectric materials can switch their polarization in response to an applied electric field. In this work, ferroelectricity at the single molecule level is proposed and investigated using density functional theory (DFT) calculations. Several bowl-shaped molecules, both synthetically reported and hypothetically proposed, are shown to invert polarization in response to external applied electric fields. Such a polarization inversion relies on the conformational change of a single molecule, unlike its traditional counterparts of which ferroelectricity originates from the switch of an asymmetrical polar unit cell in inorganic crystals, or from the polar polymer chain rotation of ferroelectric polymers. We discuss both structural and functional group factors in determining the inversion electric field and the design rules for good single molecule ferroelectrics. A conceptual multistate ferroelectric model is discussed for single molecule ferroelectrics.

Ferroelectrics are intrinsic polar materials of which polarization can be reversed by an applied external electric field. As such, ferroelectric materials have a broad range of applications including, non-volatile memory,[1-2] multiferroics,[3] and capacitor and charge storage devices.[4] The polarization inversion can be realized following different mechanisms. For inorganic ceramic ferroelectrics,[5-6] the polarization inverts accompanying the relative displacement of ions in a non-centrosymmetric unit cell. Ferroelectric polymers can perform an electric field driven polymer chain orient-reorient thus invert its polarization.[7] Dynamic proton transfer is believed to account for the ferroelectricity of some organic solids.[8-9] Novel ferroelectrics have included electron transfer in supramolecular networks[10] and orientation inversion of columnarly assembled liquid crystals of umbrella-shaped molecules.[11]

Many ferroelectrics are simultaneously piezoelectrics, which change shape in response to applied electric fields or produce electrical charge in response to mechanical distortions. We have recently reported a computational exploration of single-molecule piezoelectrics.[12] Some spring-like polar molecules can potentially serve as an independent piezoelectric unit, unlike conventional piezoelectric materials in which the piezoelectric effect results from the alignment of each polar unit/domain.

While a polar molecular spring is not likely to maintain its deformation, as well as its electric field induced polarization change, after the applied electric field is abolished; in this work, we now suggest single molecule ferroelectrics as polar molecular bowls, which can perform a polarization inversion via a bowl-to-bowl inversion under applied electric field. In particular, we focus on a group of curved molecules called buckybowls[13-14] and their hypothetical derivatives (**Scheme 1**).

We will show computationally that with careful structure design, such molecular bowls can have a wide range of inversion field, large spontaneous polarizations, and a large piezoelectric coefficient. Such single molecule ferroelectrics are intrinsically small and may be good candidates for future data storage applications.

**Computational Methods**

We used Gaussian 09[15] and density functional theory (DFT), with the B3LYP functional[16-17] and the 6-31G(d) basis set to optimize all computed structures. Although DFT methods are known to be asymptotically incorrect and ignore dispersion,[18-24] the trend of the geometry optimization on similar molecules has be shown consistent among different functionals,[12] Since no experimental data is yet available for comparison, we used the B3LYP method and discuss trends, which should be less dependent on specific functionals.

To consider the electric field induced conformation inversion and the corresponding polarization inversion, an electric field with a specified direction and magnitude was added to the Gaussian input, and the molecules were oriented to a specific frame of reference using Avagadro[29] (Supporting information, **Figure S1**). Since optimization occurs in the 3N-6 internal degrees of freedom, translations and rotations of the frame of reference were removed by select choice of the coordinates in order to examine the bowl-to-bowl inversion along the direction of the applied electric field exclusively. Transition states were set as the completely planar geometry and found via optimization in the x-y plane while enforcing planarity.

**Results and Discussion**

The electric-field induced conformational changes and bowl-to-bowl inversion were computed across a range of substituted corranulene, sumaene, and related "buckybowl" aromatic frameworks, illustrated in **Scheme 1**.

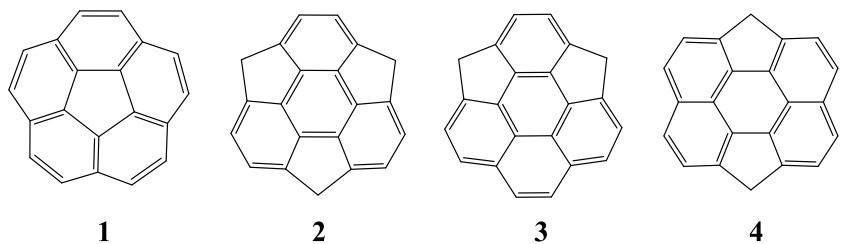

1        2        3        4

**Scheme 1.** Structures of corannulene **1**, sumanene **2**, 1,4-dihydro-as-indaceno[2,1,8,7-cdefg]pyrene **3**, 1,6-dihydrodicyclopenta[ghi,pqr]perylene **4**.

Previous studies, both computational and experimental, have shown that there is an empirical fourth-power relationship between the inversion barrier and the equilibrium bowl depth for buckybowls[30], as expressed via Equation 1, where $\Delta E$ is the inversion energy, $x_{eq}$ stands for the equilibrium bowl depth in Å, and $a$ is a coefficient tailored for a certain structure.

$$\tag{1}$$

We expect that molecules with larger inversion energy and a larger bowl depth will have a higher "inversion field",[31] if a possible electric field induced inversion would occur. We have shown in our earlier work that molecules can respond to the electric field by performing both shape change and charge redistribution.[12] Dramatic charge redistribution can possibly lead to bond dissociation and to the molecule breakdown. Therefore, it is highly possible that molecules with very high inversion energy will break down under a large applied field, rather than invert conformation.

The electric field induced bown inversion was first computed for corannulene **1** with a bowl depth of 0.86 Å. The inversion energy of corannulene is calculated to be 36 kJ/mol, close to the literature value using the same method, and to the experimental value $\Delta E$ = 42.8 kJ/mol.[32] An electric field range of ±10.28 V/nm was applied along the corannulene $C_5$ axis from the bottom to the rim of the bowl (the defined z-axis). As the field strength increases, the molecule bends inward and the bowl depth increases due to coupling between the molecular dipole moment and the applied field. The molecule is hence piezoelectric with a calculated $d_{33}$ of 13.9 pm/V within the field range of 0 - 1.29 V/nm (**Figure 1**). The trend is nonlinear over the whole range of the applied field. At a larger positive field, the nonlinear deformation is enhanced by the increased π-electron polarization. The bowl depth tends to saturate at larger negative field as increased tension will be experienced over the bowl rim due to bond length extension. Within the field range from 0 V/nm to -10.28 V/m, the bowl depth decreases only from 0.861 Å to 0.765 Å without a bowl-to-bowl inversion. An even larger electric field will not invert the bowl but deform it into a bent structure and cause a predicted C-H bond dissociation (Supporting information, **Figure S2**).

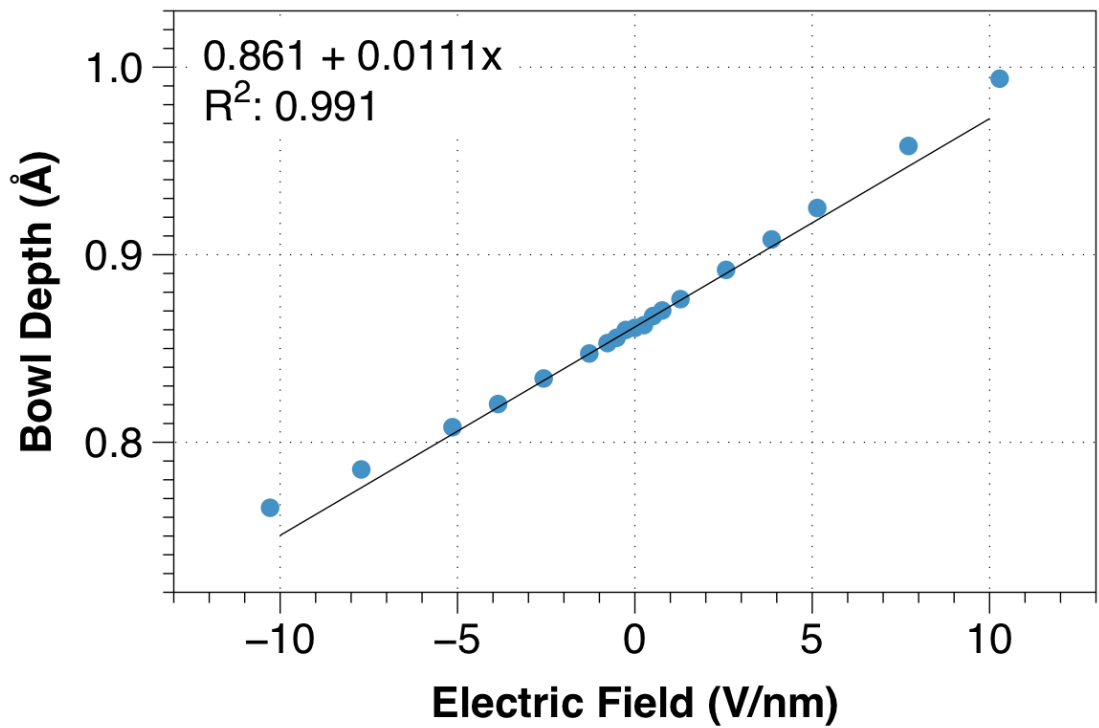

**Figure 1**. Calculated bowl depth of corannulene **1** under applied electric fields, with linear fit from -1.0 to +1 V/nm. The applied field is along the $C_5$ axis from the bottom to the rim of the bowl. A calculated piezoelectric coefficient $d_{33}$ of 13.9 pm/V is found for the linear regime.

Similar results were obtained for the stiffer sumanene **2** that no electric field induced inversion was observed over the field range of ±10.28 V/nm and a similar breakdown is assumed to take place under even stronger applied electric fields. The higher rigidity (higher inversion energy) of sumanene leads to a calculated $d_{33}$ of 5.8 pm/V over the field range of 0-1.29 V/nm, smaller than that of corannulene.

Molecules with a much smaller bowl depth must be considered for a possible field-driven bowl-to-bowl inversion. Two hypothetical molecules with a smaller bowl depth were constructed from sumanene. A shallower structure was obtained by replacing one of the five-membered ring in sumanene with a six-membered ring yielding the substituted pyrene **3**. A second structure was constructed as a substituted perylene **4**. As some of the ring strain is released by replacing a five-numbered ring with a six-membered ring, the bowl depth of **3** and **4** decreases to 0.632 Å and 0.512 Å, respectively, from 0.861 Å of **2**. The empirical quadratic relationship between inversion energy and bowl depth[30] is satisfied for the four basic structures considered (**1**-**4**) (Supporting information, **Figure S3**).

Based on the smaller bowl depth and thus smaller inversion energy, we will discuss the substituted perylene **4** first. The computed bowl depth of the optimized structure is measured as a function of the applied electric field from -3.1 to 3.1 V/nm (**Figure 2**). The positively charged bowl rim is pushed towards the bowl bottom as the electric field becomes more negative. When the field reaches -2.78 V/nm, the molecule adopts a nearly flat conformation (**Figure 2 B**). The discontinuity of the depth change indicates the bowl inversion between -2.78 V/nm and -2.83 V/nm. The inversion transition state (a flat conformation) was not captured due to the relatively large default field step (~0.05 V/nm) in Gaussian. After the inversion, the bowl depth is identical to that of the optimized bowl under the field with the same strength but in an opposite direction (e.g. -2.83V/nm to +2.83 V/nm). When the field increases, the bowl maintains its inverted direction and inverts again beyond +2.78 pm/V. Through this electromechanical distortion the calculated piezoelectric coefficient $d_{33}$ of **4** is 70.3 pm/V over the field range of 0-1.29 V/nm. The bowl depth change is more sensitive to the electric field change close to the inversion field. When field range of -2.73 to -2.57 V/nm is considered, the calculated $d_{33}$ reaches up to 600 pm/V.

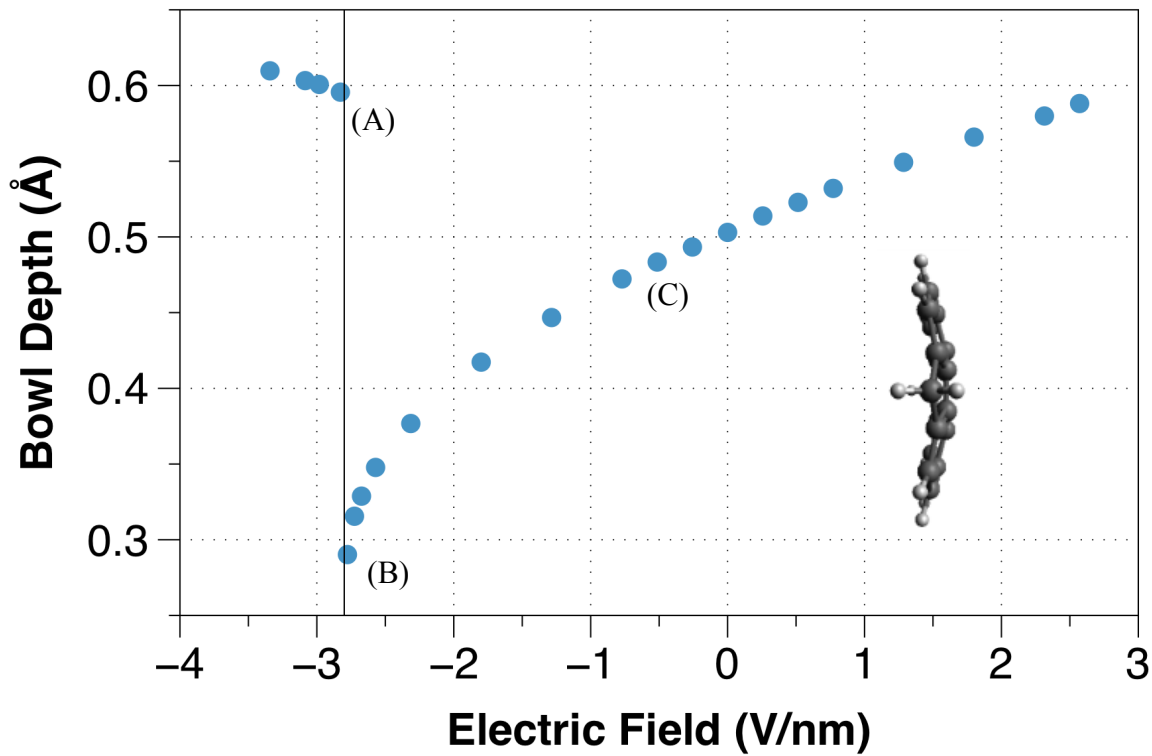

**Figure 2**. Bowl depth of perylene **4** under different applied electric field. Bowl inversion occurs at a field of -2.8 V/nm. The calculated $d_{33}$ is 70.3 pm/V across the field range of 0-1.29 V/nm.

The calculated dipole moment of **4** is 0.97 D along the z-axis, in the positive z-direction. The calculated dipole - field relationship is shown in **Figure 3**. When the electric field changes from 0 V/nm (**Figure 3 A**) to around -2.7 V/nm, the molecule dipole moment decreases almost linearly. Major dipole moment changes are found when field is approaching - 2.7 V/nm, in which the molecule is "pushed" into a relatively flat conformation (**Figure, 3 B and Figure 2 B**). The inverted perylene adopts a conformation with a much larger curvature (**Figure 3 C**) and a more negative dipole moment (i.e., opposite the original direction).

When increasing fields were applied to the inverted conformation, only gradual and minor dipole moment change was found before the field reaches up to + 2.7 V/nm. At zero field, the dipole (as well as the shape) of the inverted molecule (**Figure 3 D**) mirrors the molecule at its initial state (**Figure 3 A**). A second dipole inversion takes place between field strength of + 2.78 V/nm and + 2.83 V/nm, similar to the first inversion at the negative field range. The hysteresis feature of the dipole-field loop indicates the ferroelectric nature of perylene **4**.

The estimated spontaneous polarization without applied electric field of perylene **4** is ∼ 1.1 $\mu$C/cm$^2$. Upon functional group modifications, as discussed below, polarization of up to 4.5 $\mu$C/cm$^2$ is predicted for a cyano-substituted corannulene (**Scheme S1**, **9c**). Such high spontaneous polarization is comparable to that of inorganic (5-75 $\mu$C/cm$^2$) and organic (∼10$^{-3}$ to 13 $\mu$C/cm$^2$) ferroelectrics.[33]

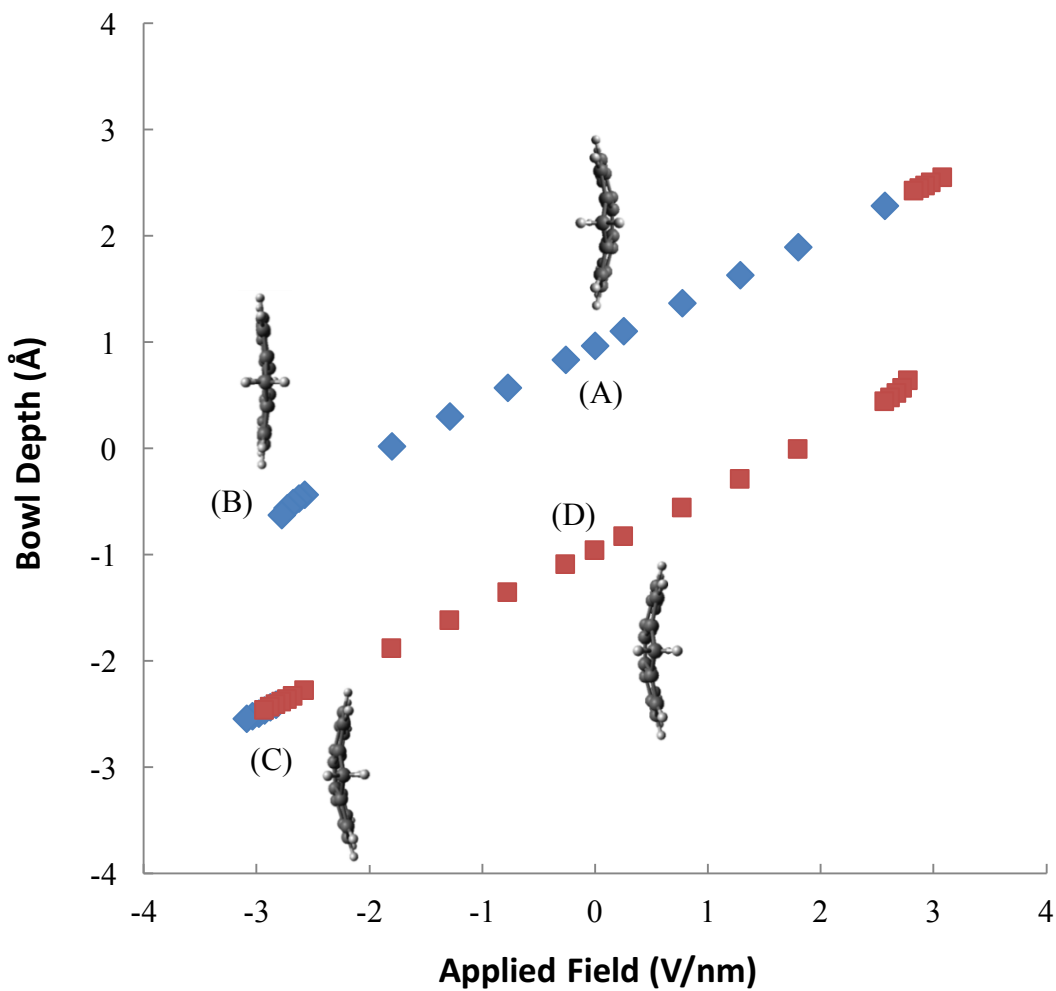

**Figure 3**. Dipole moment change of perylene **4** under different applied electric fields, indicating hysteresis effects. At zero field, conformation **A** and **D** mirror each other along the z-axis.

Computing the electric field modulation of pyrene **3**, the deeper molecular bowl, a similar bowl-to-bowl inversion was found at a higher inversion field of ∼ ± 3.9 V/nm. The dipole-field loop has a similar hysteresis feature. The calculated $d_{33}$ within the same field range (0-1.29 V/nm) of **3** is 45.0 pm/V, predictably smaller than that of perylene **4**. Such comparison indicates that buckybowls with a smaller bowl depth are more flexible, deforming more easily in response to an applied electric field. However, more structures are required for a detailed examination of the relationship between bowl depth and inversion field, as discussed in the following section.

In essence, these compounds use the potential energy of interaction between the molecular dipole moment and induced dipole (due to polarization) and the applied electric field to significantly drive the bowl-to-bowl inversion vibrational mode. Consequently, functional group substitutions that increase the z-axis dipole moment and/or decrease the bowl depth and inversion barrier will be beneficial.

Previous studies have shown that substituents on corannulene can either increase or decrease the inversion barrier depending on whether the bowl depth is increased or decreased.[30, 34] The repulsion between two *peri* substituents is likely to increase the distance between the two *peri* carbon atoms and hence decrease the bowl depth. If the two *peri* substituents tether to form a five-numbered ring (or a six-numbered ring), the distance between the two *peri* carbon atoms will decrease, giving an increased bowl depth. Several substituted corannulenes and perylenes are constructed (Supporting information, **Scheme S1**) to examine the effect of the bowl depth on the inversion field. All calculated data are listed in **Table S1** (Supporting information), including bowl depth, inversion energy, inversion field, dipole moment, and $d_{33}$.

Of the 29 structures (Supporting information, **Scheme S1**) examined, a larger bowl depth is generally accompanied with a higher inversion energy. The quadratic relationship between inversion energy and bowl depth is consistent for all corannulenes and perylenes (**Figure 4**). Different substituents provide a large range of bowl depth (0.597 Å to 1.065 Å for corannulenes and 0.332 Å to 1.005 Å for perylenes) and a broad range of inversion energy. A minor jump in bowl depth can significantly increase the inversion energy due to the quadratic relationship.

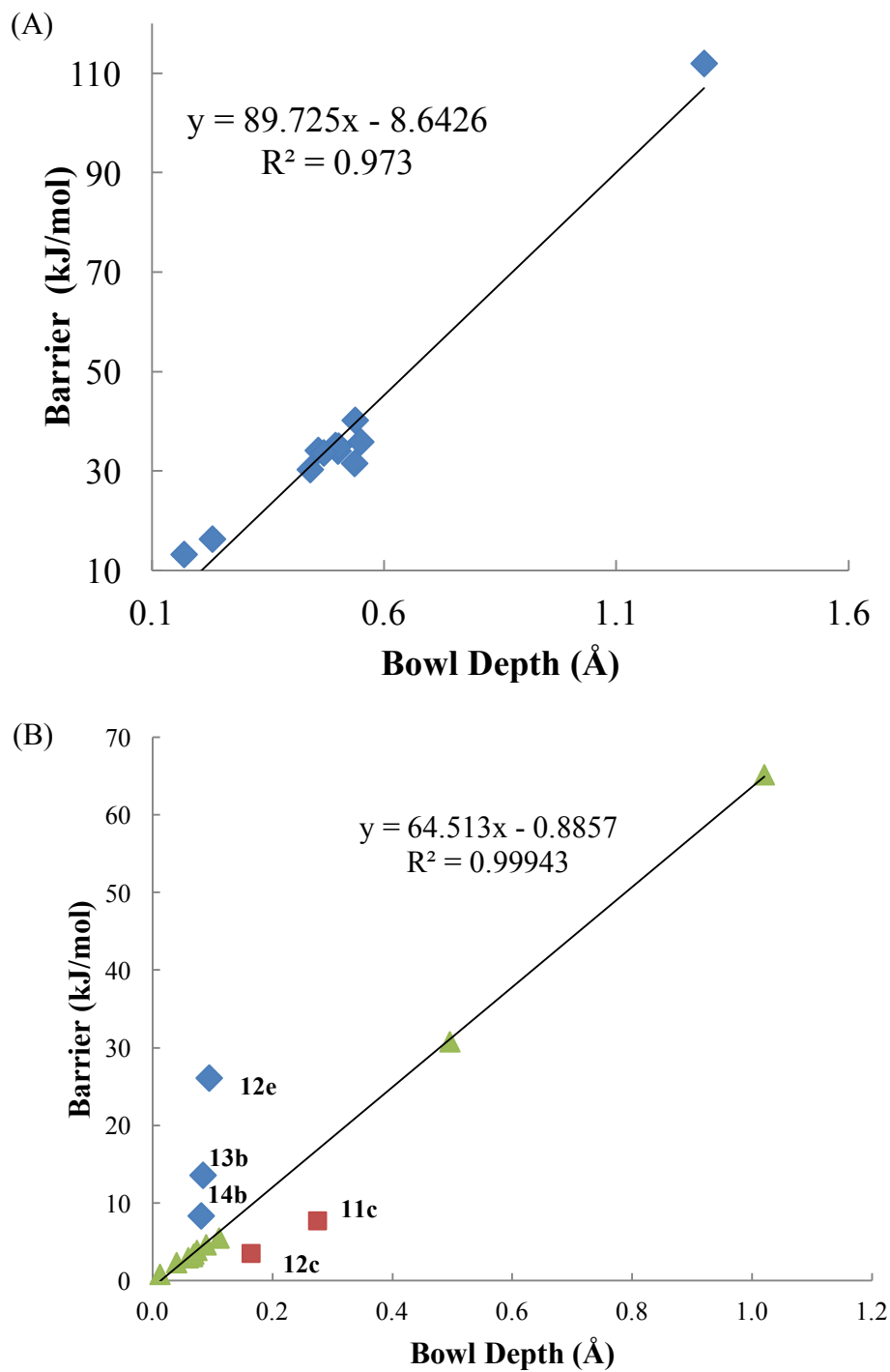

**Figure 4**. The computed inversion energy - bowl depth correlation of corannulenes (A) and perylene (B). The C-O bond rotation of -OH groups and the inversion of pyramidal -NH$_2$ complicate the calculation of bowl-to-bowl inversion.

The apparent deviations from the linear fit are the two hydroxyl- and amino- substituted perylenes (**11c, 12c, 12e, 12b, and 14b**) (**Figure 4B**, molecular structure see Supporting information **Scheme S1**). For the two hydroxyl-substituted perylenes the deviation in energy might result from the rotation of the C-O bond which could lower the energy of the transition state. In the case of the amino-substituted perylenes, the planar conformation assumes that the H-N-H angle in -$NH_2$ group is close to 120˚, and all the atoms are in the same plane. However, this may not be true when those molecules invert. The inversion barrier of pyramidal -$NH_2$ in amines is around 20 kJ/mol,[35] which can be overcome by an applied electric field as well. The inversion of pyramidal –$NH_2$ could change the molecular dipole moment dramatically and complicate the inversion of the molecular bowl. The pyramidal –$NH_2$ inversion will be discussed in detail later in this work.

Substituents play a significant role in tuning the molecule bowl depth and the inversion field. In general, shallower bowls are more likely to be inverted by an electric field (**Figure 5**). As the bowl depth decreases, an inversion field below 10 V/nm starts to appear for substituted corannulenes. A large range (~ 8.8 V/nm) of the inversion field is observed across all 29 structures with the lowest inversion field of 0.26 V/nm for octacyanoperylene (**11a**). Such low inversion field is well below the breakdown voltage of dielectric polymer PVDF (0.77 V/nm),[36] making single molecule ferroelectrics more promising in real applications.

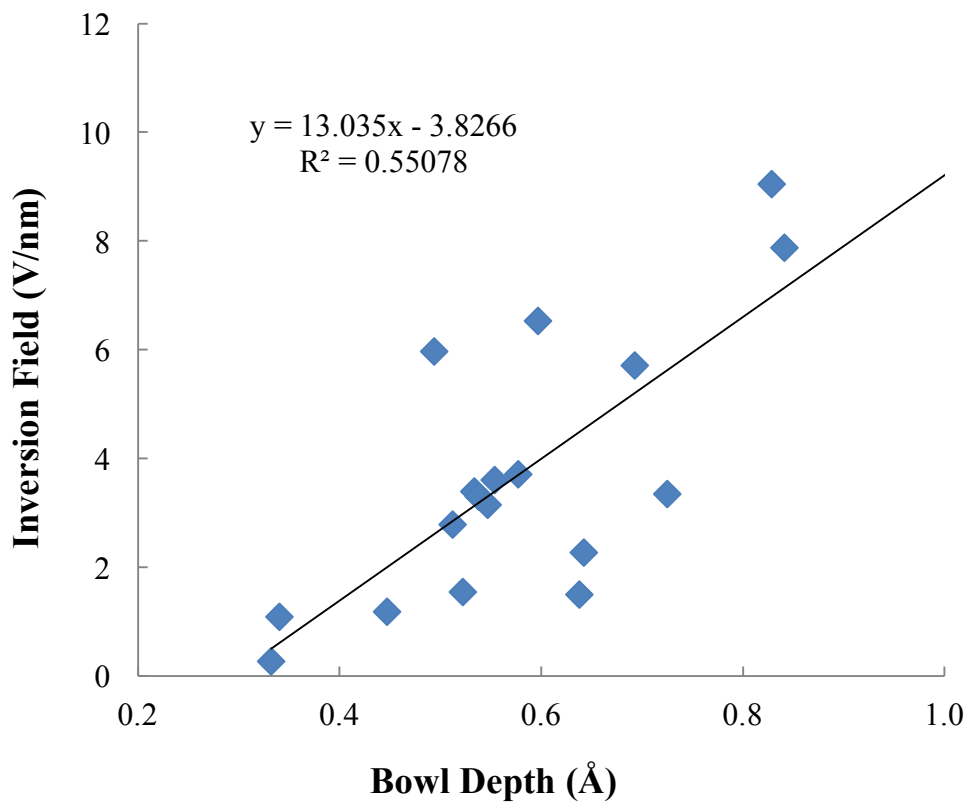

**Figure 5**. Computed inversion field - bowl depth correlation of all corannulene, perylene, and their derivatives. The inversion field is absent for some molecules within the field range considered, as listed in **Table S1** (Supporting information).

For PZT ferroelectric (ultra)thin films and alike, the coercive field *increases* as the film becomes thinner, following an empirical law $E_c(d) \propto d^{-2/3}$, where $E_c$ stands for the coercive field and $d$ for the film thickness.[37-38] The increase of the coercive field in ultrathin ferroelectric films may partially result from the effect of a conductive nonferroelectric layer between the film and the metal electrode.[39] As the film becomes thinner, the screening effect of the conductive layer (a voltage drop) reduces the voltage applied to the ferroelectric film. It is worth noting that the molecular bowl depth and PZT film thickness are different in nature and thus not a valid comparison in concept. A deeper molecular bowl stands for a higher inversion barrier and thus a higher inversion field; while a thicker PZT film doesn't change much of the inner crystal structure.

In the case of corannulenes, it seems that either a strong electron withdrawing group (-CN, -NO$_2$, etc.) or a strong electron donating group (-OH, -NH$_2$, etc.) could significantly lower the inversion field without dramatically decreasing the bowl depth. For example, pentacyanocorannulene (**10c**) has a slightly larger bowl depth (0.841 Å) than that of pentamethylcorannulene (**10b**) (0.839 Å), but **10c** inverts at a much lower electric field (7.87 V/nm) than **10b** (>10 V/nm or none). However, such trend is not so clear in the case of perylenes.

It is noticeable that for all the molecules considered (except for 1-aminoperylene, **14b**), the dipole moment is in the same direction of the inversion field (**Scheme 2**). It suggests that the electric field induced inversion is a result of the dipole-field coupling, with limited charge redistribution within the molecule π-orbitals. This also helps to explain the lower inversion field of structures with stronger electron withdrawing/donating groups. For example, tetrahydroxylperylene (**12c**) has a smaller inversion field (1.49 V/nm) than that of tetramethylperylene (**12b**) (3.14 V/nm), though the former molecular bowl is deeper (0.638 Å of **12c** to 0.547 Å of **12b**).

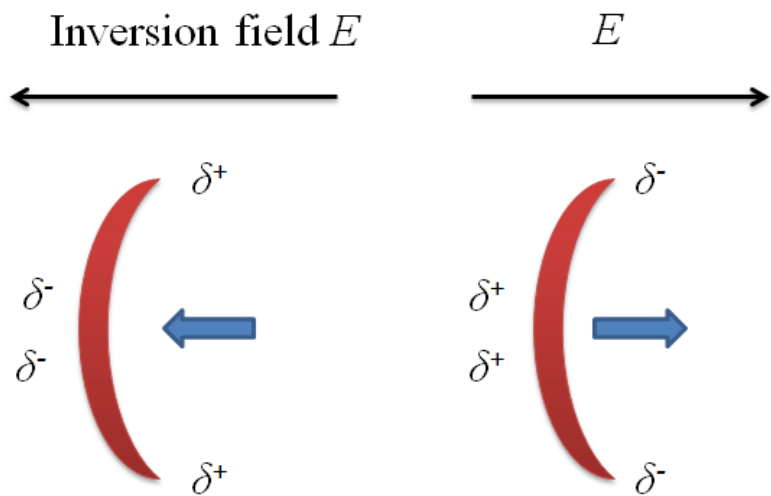

**Scheme 2**. Illustration that the inversion field is applied in the same direction of the molecule dipole moment.

The change of the bowl depth, or the molecular "thickness", to the electric field of all invertible structures is similar in shape (see **Figure 2**), only with the inversion field occurring at different strength of the electric field. A linear trend of bowl depth change is observed within a small electric field range (e.g. within ± 1.29 V/nm) and a dramatic jump close but before the inversion field. Surprisingly large piezoelectric coefficients are observed for substituted perylenes. All of the calculated $d_{33}$ values are listed in **Table S1** (Supporting information). Over the field range of 0 - 1.29 V/nm, the $d_{33}$ of octacyano-substituted perylene (**11a**) reaches up to 450 pm/V, comparable to the best conventional PZT piezoelectrics. Such a significant $d_{33}$ for **11a** may result from the small bowl depth (0.332 Å) and the resulting flexibility of the bowl, combined with the large dipole moment (2.38 D).

The trend is not clear in both the relationship of bowl depth to $d_{33}$ and the relationship of dipole moment to $d_{33}$ (Supporting information, **Figure S4**). However, the $d_{33}$ seems to decrease exponentially as the inversion field increases (**Figure 6**). Similar phenomena have been reported for conventional PZT ferroelectrics and alike.[40-42] The so-called size effect for ceramic piezoelectrics describes the decreasing trend of $d_{33}$ to the decrease of the film thickness. On the other hand, the $d_{33}$ decreases as the coercive field increases based on the fact that thinner films tend to have a higher coercive field. Such similarity, along with the significant calculated piezoelectric coefficient of perylene derivatives, may shed lights on the possible replacement of conventional perovskite ferroelectrics by their counterparts of single molecule piezoelectrics for future small dimension, flexible, high density, and patterning-rich applications.

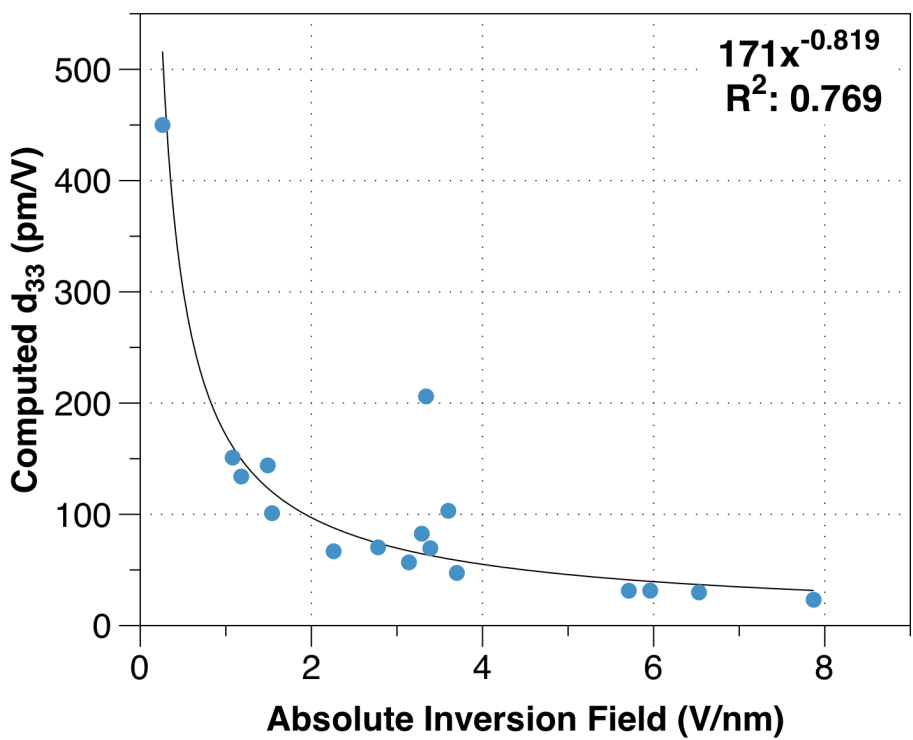

**Figure 6**. Correlation of $d_{33}$ to bowl depth and inversion field. Magnitude of the inversion field is considered as a measure of the inversion barrier.

The inversion of pyramidal -NH$_2$ groups mentioned earlier will change the molecular dipole moment dramatically without inverting the molecular bowl. Depending on the orientation of the -NH$_2$ groups, the overall dipole moment of amino-substituted buckybowls could reside at four different states of different magnitude and/or directions. Such molecules can be viewed as a model of multistate ferroelectrics at low temperature.[2, 43-44] The working principle of this multistate ferroelectric model is illustrated using the tetraamino substituted perylene (**12e**) as an example in **Figure 7**.

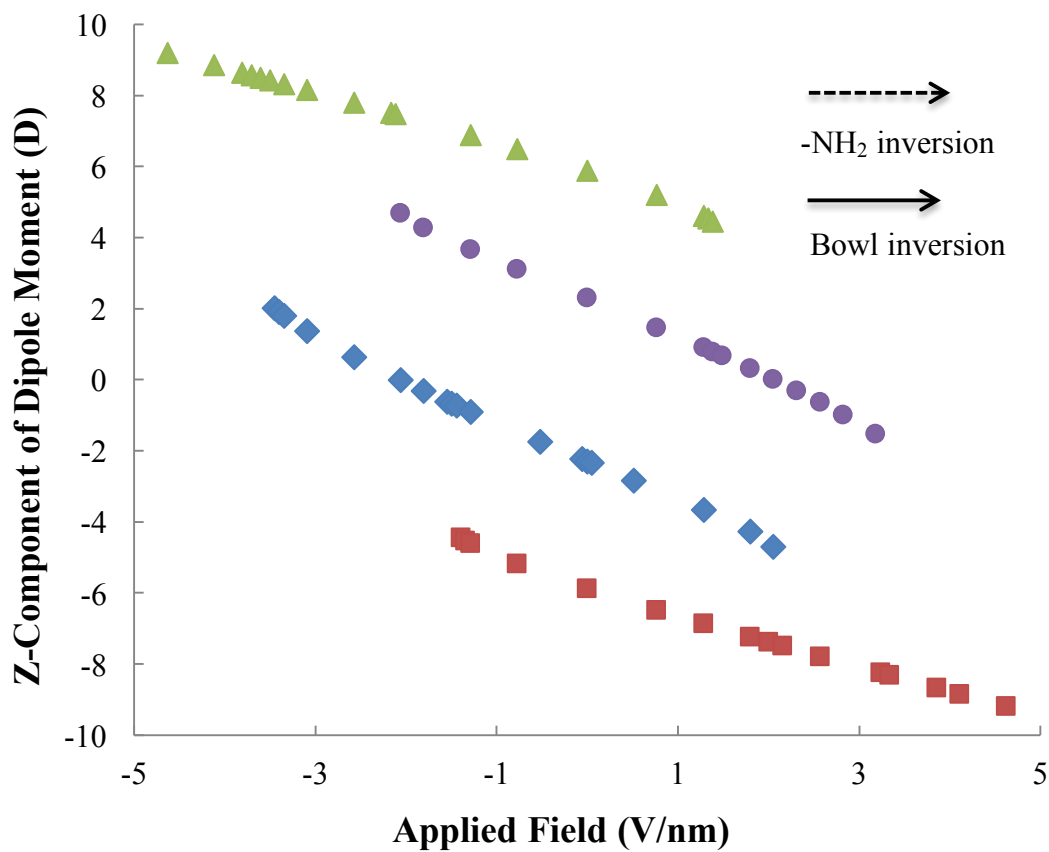

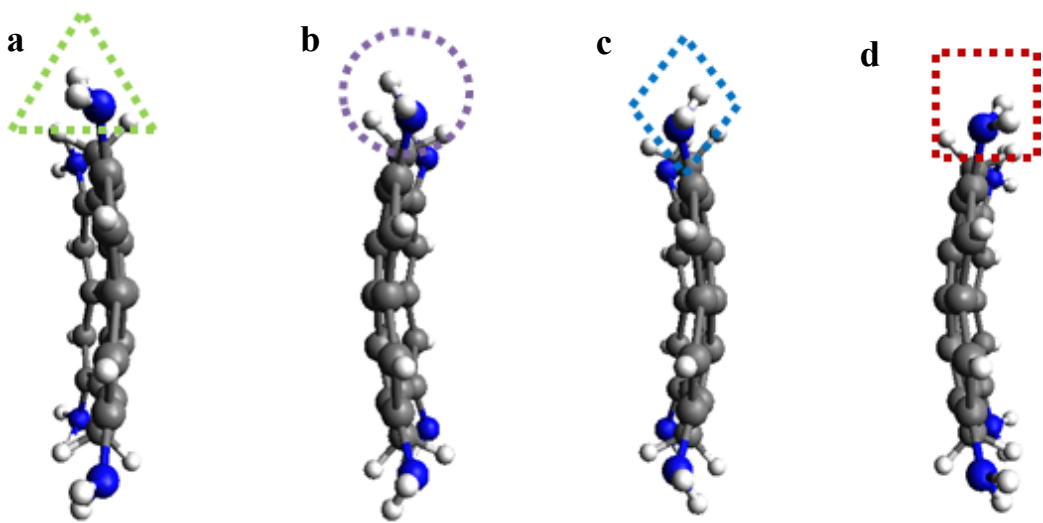

**Figure 7**. (Top) Z-component of the dipole moment of perylene **12e** under different applied electric fields. (Bottom) The four stable conformations adopted by **12e** at 0 V/nm with a dipole moment different in magnitude and/or direction. Perylene **12e** can be viewed as a single molecular model of a four-state ferroelectric at low temperature.

Perylene **12e** has a large dipole moment of 5.88 D under zero electric field, if the two hydrogen atoms in each -$NH_2$ group bend to the positive direction of the z-axis (**Figure 7 *a***). The dipole moment decreases in a nearly linear fashion as the field increases. At ~ 1.39 V/nm the molecular bowl inverts and the dipole moment drops from 4.45 to 0.73 D. The inverted **12e** has a dipole moment of 2.29 D if zero field is applied (**Figure 7 *b***). Under larger electric field, the dipole moment continues to decrease and dramatically drops from -1.54 to -8.26 D at around 3.19 V/nm where the four -$NH_2$ groups invert to the negative direction of the z-axis (**Figure 7 *d***). When the field starts to decrease, conformation ***d*** maintains (with gradual attenuation of the bowl depth), and the bowl inverts again at -1.39 V/nm with the dipole moment increase from -4.45 to -0.73 D. The inverted conformation has a dipole moment of -2.29 D free of applied electric field (**Figure 7 *c***). The dipole moment - electric field curve is centrosymmetric with multiple hysteresis loops.

Compound **12e** has four spontaneous polarization states which can be altered by applied electric fields. Direct switching is infeasible between conformation ***a***, ***d***, and ***b***, ***c***, respectively. Switching between any two of the four states can be achieved by applying a maximum of two sequential applied electric fields.

Note that the DFT calculations assume a temperature of 0 K and thus zero vibrational energy of the molecule. So the calculated inversion field cannot be directly extrapolated to molecules at room temperature. The inversion of corannulene is observed to take place above -64°C in solution, and is estimated to take place over 200,000 times per second at room temperature.[45] The pyramidal amines with a smaller inversion barrier, 20 kJ/mol compared to to 36 kJ/mol of corannulene, will start to invert at an even lower temperature. However, giving that a huge range of inversion energy can be potentially achieved through structural and functional modification of buckybowls, it is highly promising that careful computational design and versatile synthetic strategy could lead to room temperature or even high temperature single molecule ferroelectrics.

Similarly, the multistate ferroelectricity of perylene **12e** could only be possibly achieved at very low temperatures due to the small energy barrier of pyramidal $-NH_2$. However, the concept of combining two different types of field-induced polarization inversions can still inspire the design of new multistate ferroelectrics.

On the other hand, molecular packing constraints in crystals or multilayers may prevent the molecule from inverting and elevate the inversion field. Such constraints may also obstruct the inversion of the pyramidal $-NH_2$ group hence chances of the inversion may be left for applying an electric field. Single molecule based ferroelectrics with a broad range of inversion field in room temperature or higher could be promisingly constructed combining both structures/substituents effect and molecular packing constraints.

**Conclusions**

We have computationally demonstrated the potential polarization hysteresis of single molecules under applied electric fields. The polarization inversion is achieved via molecular conformational change, namely the bowl-to-bowl inversion, of a group of hypothetical buckybowls and their derivatives. Such mechanism is different from both "displacive" and "proton-transfer" mechanisms of conventional inorganic and organic ferroelectrics. Recently, a similar idea has been experimentally realized via a hexagonal columnar liquid crystal assembly of umbrella-shaped molecules,[11] analogous in shape to buckybowls.

Both molecule structure (bowl depth) and substituents play an important role in determining the ferroelectric property of the molecules. A large bowl depth elevates the inversion energy barrier and leads to a higher inversion field. Molecules with a barrier larger than 34 kJ/mol are likely to break down rather than invert under strong electric field. The bowl depth also changes as different substituents are introduced to add or release tension of the curved bowl. For example, the bowl depth of corannulene is likely to decrease due to the repulsion from two *peri* substituents and to increase if the two *peri* substituents bridge to a five-numbered ring. Both the magnitude and direction of the dipole moment can be changed by different substituents which change the polarization and the sign of the inversion field.

We find many potential single molecule ferroelectrics, with a large range of inversion field (~8.8 V/nm), a high spontaneous polarizations (up to 4.5 $\mu$C/cm$^2$ for **9c**), and a huge piezoelectric coefficient $d_{33}$ (up to 450 pm/V for **11a**). Such values are comparable or even superior to all conventional ferroelectrics. Using the diversity and tailorability of organic chemistry, a new class of molecular ferroelectrics based on conformational changes may be created.

AUTHOR INFORMATION

**Corresponding Author**

*Email: geoff@pitt.edu

The authors declare no competing financial interests.

ACKNOWLEDGMENT. This research was performed, thanks to the donors of the American Chemical Society Petroleum Research Fund (49002-DN110), AFOSR (FA9550-12-1-0228), and RCSA through the Cottrell Scholar award for funding.

**Table of Contents Graphic:**

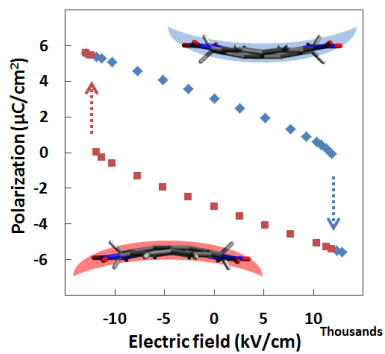